\documentclass[11pt,twoside]{article}
\usepackage{asp2014}
\aspSuppressVolSlug
\resetcounters

\begin{document}

\title{Revealing the Galactic Population of BHs}
\author{Thomas J. Maccarone,$^1$ Laura~Chomiuk,$^2$ Jay Strader,$^2$ James Miller-Jones,$^3$ Greg Sivakoff$^{4}$
\affil{$^1$Texas Tech University, Lubbock, TX, USA; \email{Thomas.Maccarone@ttu.edu}}
\affil{$^2$Michigan State University, East Lansing, MI, USA; \email{chomiuk@pa.msu.edu}, \email{strader@pa.msu.edu}}
\affil{$^3$Curtin University, Perth, WA, Australia \email{james.miller-jones@curtin.edu.au}}
\affil{$^4$University of Alberta, Edmonton, Alberta, Canada \email{sivakoff@ualberta.ca}}
}

\paperauthor{Sample~Author1}{Author1Email@email.edu}{ORCID_Or_Blank}{Author1 Institution}{Author1 Department}{City}{State/Province}{Postal Code}{Country}
\paperauthor{Sample~Author2}{Author2Email@email.edu}{ORCID_Or_Blank}{Author2 Institution}{Author2 Department}{City}{State/Province}{Postal Code}{Country}
\paperauthor{Sample~Author3}{Author3Email@email.edu}{ORCID_Or_Blank}{Author3 Institution}{Author3 Department}{City}{State/Province}{Postal Code}{Country}

\begin{abstract}
  We discuss the case for using the Next Generation Very Large Array
  both to discover new black hole X-ray binaries astrometrically, and
  to characterize them.  We anticipate that the ngVLA will be able to
  find $\sim$100 new black hole X-ray binaries, as well as a host of
  other interesting radio stars, in a few hundred hour survey.
  Parallax and astrometric wobble measurements will be achievable in
  feasible follow-up surveys especially using long baseline
  capabilities.  The ngVLA's high angular resolution, high survey
  speed, and high frequency sensitivity give it a unique range of
  parameter space over which it is sensitive.
\end{abstract}

\section{Description of the problem}

The ngVLA will illuminate the formation processes of BHs (BHs) in close binaries through high angular resolution, high sensitivity radio
observations. A multi-epoch Galactic survey with the ngVLA will answer key questions like:
 How many
binaries host BHs?  What kinds of cosmic cataclysms produce
BHs? How do BHs end up in close systems with other
compact objects, in order to merge and produce gravitational waves?

Such a survey is timely, as we currently stand at the dawn of gravitational wave astrophysics. In recent years, we have seen the discoveries of stellar-mass BH-BH binaries merging, as well as one double neutron star merger (Abbott et al.\ 2016, 2017) . While LIGO demonstrates the existence of inspiraling BH binaries (and can measures key parameters like mass and spin for
these objects), understanding the origin of these systems requires information that can only come from electromagnetic studies of BH binary progenitors.  Key questions remain about whether
these objects form via standard binary stellar evolution
(e.g. Belczynski et al. 2007), triple star evolution (e.g., Rodriguez \& Antonini 2018), chemically homogeneous binary evolution
(de Mink et al. 2009), or dynamically in globular clusters
(e.g. Miller \& Hamilton 2002). The rates of these mergers are
dependent on a variety of poorly constrained parameters, including,
but not limited to common envelope efficiency, BH natal kicks
distributions, and the range of masses of BHs in binaries.
Understanding the populations of BHs in binaries in our own
Galaxy will yield crucial constraints on all of these.

Furthermore, the same data that will yield the measurements needed to
understand double compact object mergers will give vital insights into
how the supernovae themselves explode.  
We currently know of only $\approx$20 dynamically confirmed stellar-mass BHs in
our Milky Way. The known objects have mostly been discovered via bright X-ray outbursts, and then followed up spectroscopically in quiescence. This selection function creates a strong bias toward systems with long orbital periods (e.g. Arur \& Maccarone 2018)
Building a sample of accreting BHs in quiescence through a
sensitive radio survey will yield a sample with much better understood selection effects. 
The ngVLA can build such a sample, with suffiicient sensitivity to detect quiescent BH X-ray binaries over a large volume of our Galaxy, and sufficient resolution to measure proper motions and thereby filter out background extragalactic contaminants.


\section{Scientific importance and Astronomical Impact}

The total number of accreting stellar-mass BHs in our Galaxy is wildly
unconstrained, potentially ranging from $100-10^8$ (Tetarenko et
al. 2016), with clear dynamical signatures of about 20 (Casares \&
Jonker 2014) and about 60 strong candidates (Corral-Santana et
al. 2016).  Arguments based on the metal enrichment of
the Galaxy suggest that there should be $\sim 10^8-10^9$ stellar-mass
BHs in the Galaxy (e.g. Samland 1998), and this
number is consistent with the upper limits from microlensing searches
(e.g. Wyrzykowski et al. 2016).  Typically 1--2 new stellar-mass black
holes in outburst are discovered per year, and almost all of the known
stellar-mass BHs are on the near side of the Milky Way Galaxy.
These facts, plus modelling of selection effects suggest that the total number of BH low-mass X-ray binaries is likely to be a few thousand or more (Arur \& Maccarone 2018).

The number with high mass donor stars is even more uncertain, in part
because wind-fed systems may evade discovery because they do not host
X-ray outbursts (see e.g Tetarenko et al. 2016).
There may, in fact, be a population of systems very similar to the canonical high mass X-ray binary BH system Cygnus X-1, except with wider separations so that they have lower accretion rates and X-ray luminosities. These systems are poor bets to be discovered in X-ray and optical/IR surveys because of the broad similarities between their expected appearance and the expected appearance of generic massive stars. However, they should be readily distinguished at radio wavelengths by their spectral index; the flat radio spectra produced by an accreting BH's jets will contrast with the optically-thick thermal spectrum expected for a massive star:

An improved understanding of the population size of accreting BHs in
the Milky Way will place strong constraints on some of the largest
open questions in stellar astrophysics today. It will be a crucial
constraint on the efficiency of common envelope evolution in close
binary systems (e.g. Ivanova \& Chaichenets 2011). It will also
constrain the strength of kicks granted to BHs in their natal
supernovae.

The amplitude of BH kicks can illuminate the primary channel of
stellar-mass BH formation. It has been speculated that BHs may form
from a prompt collapse of the entire massive star at the end if its
lifetime or, alternatively, from fallback accretion on to a neutron
star. The latter case should apply a much stronger natal kick to the
BH at the time of formation, both because of the temporary presence of
the neutron star (e.g. Kalogera 1996) and the symmetric mass loss from
a moving object in a binary (Blaauw 1961).  Measurement of BH kicks is
also important for determining if LIGO BH binaries are produced
through normal binary evolution or dynamical interaction (i.e., in
dense star clusters). The gravitational waveform reveals the
misalignment of spin---of the black holes relative to one another, and
relative to the orbital plane. Dynamically formed binaries should have
an approximately random distribution of spin orientations, while
binaries formed through binary evolution should have some preference
for aligned spins---unless very large kicks take place (REF). It is
thus vital to determine whether typical BH kicks are large enough to
misalign BH spins substantially from their orbital planes in order to
understand how robustly misaligned spins indicate a globular cluster
formation scenario.  The scale heights of black holes and neutron
stars in current samples seem to be quite similar (Repetto \& Nelemans
2015)...
However, there does seem to be at least one case, Cygnus X-1, where the applied kick appears quite small (Wong et al. 2012). 
An ngVLA survey for accreting BHs will constrain BH kicks in two ways---by measuring the number of BHs in close binaries in our Galaxy, and by directly measuring their proper motions.


We can also place constraints on the formation of LIGO BH mergers by
comparing the BH populations in the Galactic field with those in dense
stellar environments (globular clusters, Galactic center), and
estimating the relative importance of binary and dynamical channels
for forming close BH binaries.  At the present time, there are some
strong radio-selected BH candidates in globular clusters
(e.g. Strader et al. 2012; Chomiuk et al. 2013).  Both new
detections and confirmations of existing candidates' membership can be
made via astrometry.

A larger, more representative sample of black holes will enable a high-quality measurement of the BH mass distribution, and thereby constrain some of the most poorly understood aspects of supernova explosions. 
In particular for core-collapse supernovae, there is still
no consensus about why the explosions actually take place, with the
leading models being a standing accretion shock instability and the
Rayleigh-Taylor instability (e.g. Belczynski et al. 2012 and
references within).  Because the process takes place while a thick
envelope covers the stellar core, and the supernova light curves and
spectra are largely sensitive only to the total energy input and the
mass of the envelope, other approaches to understanding the events are
vital.  In principle, gravitational waves and neutrinos provide
information at the time of explosion, but these are only detectable at
the time of explosion and within small horizon volumes. The compact remnants of supernovae are therefore the best observational clues as to the processes driving supernovae---and specifically, the distribution of the masses of the compact stellar remnants is very information-rich.
Belczynski et al. (2012) argue that models in which
the explosion proceeds on a timescale of less than about 0.2 seconds
after the core collapse will produce a substantial gap in masses
between the heaviest neutron stars and the lightest BHs, while
explosions which happen on timescales of order 0.5 seconds or more
will lead to a continuous distribution of compact object masses.  At
the present time, there appears to be such a gap when the mass
distribution is modelled as the sum of Gaussians (\"Ozel et al. 2010;
Farr et al. 2011), but other functional forms for the mass
distribution allow a continuous range of masses (Farr et al. 2011).

The present size of the BH mass sample is simply not
sufficient to establish firmly whether there is a mass gap.
Furthermore, some biases in our estimates of the inclination angles of
the binaries may be partially or fully responsible for the apparent
presence of this mass gap (Kreidberg et al. 2012).  Astrometric wobble
measurements of even a few binaries would provide gold-standard
calibration for ellipsoidal modulation measurements, and would be
possible with high sensitivity, high frequency VLBI measurements (see
also the article by Reid \& Loinard in this volume). With the identification of a substantial population of accreting BHs,
we will have in hand an ideal sample for follow-up observations enabled by both the ngVLA itself along with other multi-wavelength facilities available in the next decades.
This larger sample will also be unbiased, sampling the full population of accreting stellar-mass black holes to enable a measurement of the BH mass function and the amplitude of BH natal kicks.


 \section{Anticipated results}  
 
 The ngVLA can detect accreting stellar-mass BHs over a large volume
 of our Galaxy, and distinguish them from interlopers. An astrometric
 survey would quickly yield the ability to separate background AGN
 from foreground X-ray binaries while simultaneously giving good
 measurements of their proper motions. 
 Numerous other classes of Galactic sources will also be present in a deep, wide astrometric survey, and the combination
of the radio properties with other multi-wavelength data sets will not only allow identification of which objects are stellar-mass BHs in binaries, but also will allow characterization of these other populations. We expect to detect bright flare stars, cataclysmic variables, planetary nebulae, and pulsars (most of which should identifiable based on radio spectra, radio variability, angular extent of
radio emission or obvious associations with bright foreground stars), and also neutron star X-ray binaries, transitional millisecond pulsars and colliding wind binaries (which may require more careful follow-up work).
  A particularly interesting class of
 other objects that could be found in such a survey is isolated black
 holes accreting from the interstellar medium (Maccarone 2005; Fender,
 Maccarone \& Heywood 2013).

 The radio survey could be conducted in a reasonable exposure time,
 since objects like the radio-faintest known X-ray binaries
 (e.g. A0620-00 \& XTE~J1118+480 -- Gallo et al. 2006; Gallo et
 al. 2014) could be detected in about 6 minutes at 6$\sigma$ at a
 distance of 2-4 kpc with the proposed ngVLA sensitivity.  BHs
 accreting at higher rates, like V404~Cyg in quiescence, are a factor
 of $\sim15$ more luminous (Miller-Jones et al. 2009) and can be
 observed well past the distance of the Galactic Center.  We thus
 expect to be able to detect $\sim10\%$ of the short period BH
 X-ray binaries in a 10 square degree survey region near the Galactic
 Center. This region contains about 10\% of the stellar mass of the
 Galaxy, meaning that we would expect to detect at least 1\% of the
 total number of short period BH X-ray binaries in the Milky
 Way, and the majority of the long period, V404~Cyg-like BH
 X-ray binaries.  We thus expect that $\sim 100$ new BH X-ray
 binaries should be discovered in this proposed survey, based on
 population estimates from X-ray studies (e.g. Arur \& Maccarone
 2018), and perhaps much larger numbers based on the surprising
 discovery of a single strong BH candidate with a parallax
 placing it in front of M15 (Tetarenko et al. 2016).  We emphasize
 further that this proposed program would require a modest amount of
 time, ado a broad range of additional science, and be easily
 extendable over time both by extending the field of view and making deeper
 observations.

 At a distance of 8 kpc, with 10 mas resolution (requiring significant
 collecting area on baselines of at least 1000 km), an uncertainty on
 proper motion of 20 km/s would result from a nominal two-year time
 baseline for $6\sigma$ detections.  Thus the uncertainty will be much
 less than the typical stellar velocity dispersion in the Solar
 neighborhood, so only a small fraction of objects should be expected
 to have small enough proper motions to be confused with background
 AGN, and this fraction should decrease further if the majority of
 systems form with strong natal kicks.  

 Having long baselines (of approximately 1000 km or more) is thus
 crucial in order to allow sufficiently precise proper motion
 measurements to make the proper motions diagnostics sufficient both
 for ruling out background AGN sources, and allowing the measurements
 to yield useful information about natal kicks.  While the proper
 motion uncertainty increases for sources behind the Galactic Center,
 if such sources follow the Galactic rotation curve, they will
 have space velocities of about 400 km/sec because of the rotation
 relative to the Earth's motion, so even sources at such distances, if
 bright enough to be detectable, should show measureable proper
 motions, even in the absence of natal kicks.


\section{Limitations of current astronomical instrumentation}

Current long baseline arrays are simply not sensitive enough to meet
our science goals. It would require 23 hours per pointing to reach our
per-epoch sensitivity goal with the LBO---a factor of 4000 longer than
with the ngVLA, and to cover 10 square degrees at 6 GHz would require
about 640 pointings, so that it would take about 7 years at 6 hours
per day to conduct this survey.  The European long baseline facilities
are too far north to do this work effectively (e.g. e-Merlin sees the
Galactic Center region as a maximum elevation of 7 degrees).  The VLBA
also runs into some problems with source scattering in the most
scattered parts of the sky so that only a small fraction of its
baselines become useful in such regions.

While the \emph{Gaia} mission holds some promise for identifying BHs
in very wide binaries from their astrometric wobble (Barstow et
al. 2014; Mashian \& Loeb 2017), it will struggle to obtain results on
the optically faint quiescent binaries in the crowded,
dust-extinguished region of the Galactic Bulge.  This dense and
populous region is essential for understanding the Galactic BH
population and how its characteristics depend on density.
Furthermore, the orbital period range probed by \emph{Gaia} through
astrometric wobble is much longer than the orbital period range probed
by detecting accreting systems -- if \emph{Gaia} does discover many
stellar mass BHs as wobblers, this information is complementary to
what radio observations could discover.  Only about 10\% of the strong
candidate BH X-ray binaries known are bright enough optically for
\emph{Gaia} to detect them, despite the strong biases in existing
samples toward nearby, unreddened objects.

\section{Connection to unique ngVLA capabilities}

The long baselines ($\sim$1000 km) and high resolution of the ngVLA
are critical to this astrometric survey. In addition, relatively high
frequencies ($>$ 4 GHz) are necessary to obtain good image quality in
the Galactic Plane and the Galactic Bulge, as low frequencies will be
affected by scattering.  A broad range of baselines is needed for
these studies in order that even in the most confused regions, like
those in Cygnus and those very close to the Galactic Center, good
angular resolution is possible without over-resolving heavily
scatter-broadened sources.

\subsection{Benefits to this science from a more extended configuration}

In addition, continental and global baselines would be of
tremendous interest for follow-up of many of the sources discovered
through an astrometric survey.  Most of the known X-ray binaries are
too faint optically for \emph{Gaia} to provide parallax distances.  Absolute
astrometry with the \emph{James Webb Space Telescope} is expected to be
limited to about 0.5 milliarcsecond, even at arbitrarily large signal
to noise, meaning that geometric parallax distances will be limited to
about 1 kiloparsec.  As a result, the best geometric parallax
distances for these sources will come from radio data, rather than
from optical data.  Follow-up could be done with only antennae on the
outer part of the array, plus either VLBA stations, or newer longer
baseline ngVLA dishes.  While the follow-up would require more time
per target, it could be done at higher frequencies to improve
positional accuracy in a given integration time and mitigate the
effects of scattering, and it could be done only for the sources of
interest.  It would make an excellent usage of the portion of the
array that would not be needed when most of the core antennae are
being used for low surface brightness extended source projects.  To do
this follow-up though, a sample of X-ray binaries with known radio
brightness must first be collected.

\section{Experimental layout}

We plan to observe in the band covering 4--12 GHz, as a compromise
between field of view, resolution, and sensitivity (we expect the
spectra of accreting BHs to be flat, $S_{\nu} \propto \nu^0$). We
estimate that the required sensitivity of each epoch of the survey is
0.8 $\mu$Jy beam$^{-1}$ (achievable in 6 minutes on source). We require
at least three visits covering the survey area, in order to obtain
high-quality proper motion measurements. We anticipate achieving an
image resolution of $\sim$10 mas by working at medium frequency with
1000 km baselines.

By covering 10 sq. deg. of the Galactic bulge, we will survey the BH
population over a substantial fraction of the Galaxy and span a range
of environments, including the dense central regions and star-forming
regions and the Galactic field. We estimate that each epoch will
require 100 hours on source, implying that the entire survey can be
carried out in 300 hours.

After doing preliminary classification of the radio sources based
solely on their radio properties, the AGN, pulsars, planetary nebulae,
cataclysmic variables and isolated massive stars should
have been filtered out of the data sample.

Many of these other classes of Galactic stellar radio sources will be
interesting in their own rights.  Remaining after filtering will be
various classes of BH and neutron star X-ray binaries, active binaries
and colliding wind binaries.  For these, multi-wavelength follow-ups
will be necessary.  The active binaries will be the most challenging class of contaminant of the X-ray binary population, but the bulk of these should be identifiable from circular polarization for relatively high signal-to-noise sources.  For fainter sources radio/X-ray ratios, which will be high enough to be X-ray binaries only in flaring states, will be useful, since the source are highly unlikely to be flaring in all epochs.

\section{Complementarity with searches at other wavelengths}

Non-simultaneous coordination with the WFIRST microlensing planet
search fields will provide high cadence infrared data and thereby
ellipsoidal modulation measurements to estimate orbital periods and
inclination angles for many of the objects.\footnote{Because of the
  very large number of W UMa stars, using the WFIRST data alone is
  unlikely to produce good catalogs of BH and neutron star
  candidates.} BH mass estimates can be obtained using optical/IR
spectroscopy to measure the width of emission lines, in conjunction
with the Casares (2016) relation. A healthy diverse portfolio of
optical/IR telescopes spanning diameters, 4--30m, would be ideal for
the spectroscopic follow-up, but it is likely that once the sources of
interest are identified that most should be bright enough that the
Casares relation or some infrared equivalent will be suitable for
estimating the radial velocity amplitude from emission lines.

Proposed future X-ray missions such as eROSITA and Athena (which are
scheduled to proceed), and Lynx and STAR-X can be expected to deliver
substantial populations of faint candidate X-ray binaries,
interspersed with large numbers of members of other X-ray source
classes.  Because large fractions of these will be radio sources as
well, particularly at the sensitivity levels we discuss here, the
relatively uncrowded radio data, which will carry with them proper
motion infomation, will help sort out the identification of these sources.

\section{Additional astrometric stellar-mass BH science}
Apart from the value of a large, dedicated Galactic survey, the ngVLA,
especially with long baselines, would be vital for a variety of other
stellar-mass BH science.  In particular, with $\mu$Jy
sensitivity on 5000 km baselines, at 40 GHz would allow for geometric
parallax distance measurements for most of these same stellar-mass
BHs the survey would find, and for many accreting neutron
stars, even in heavily scattered regions.  Doing this requires, at
bare minimum, retaining the current VLBA stations and upgrading their
receivers to make them part of the ngVLA, but exposure times a factor
of 5 shorter can be obtained for the same precision by putting about
10\% of the collecting area on baselines longer than 1000 km.

We can anticipate that there would be about 300 baselines from the New
Mexico core to the 10 VLBA antennae, so that the sensitivity would be
about 5.5 times worse than the ngVLA's overall sensitivity for VLBI
experiments.  Putting 10\% of the collecting area on longer baselines
would improve the sensitivity by a factor of about 2.6, reducing the
exposure time needed for such projects by a factor of about 5, and
making the long baseline sensitivity only about a factor of 2 worse
than the overall sensitivity.

In the case with no new collecting area beyond 1000 km, follow-up
measurements to obtain parallax distances would require about 30 times
the total exposure needed to detect the sources, and five epochs would
be needed to make reliable parallax measurements.  Since the survey
observations are expected to be only 6 minutes each, this would
require 15 hours per object, so that if 100 BHs were detected,
this would become a 1500 hour project at most.  In practice, many of
the source will be significantly brighter than the detection limit,
and some of the foreground objects may have detections from Gaia as
well.  If 10\% of the collecting area were on long baselines, this
survey could be done in about 1/5 the time, or, in approximately the
same amount of time using the long baselines alone, and with somewhat
better astrometric precision since a larger fraction of the collecting
area would be at baselines longer than the baselines to New Mexico.
With 0.2 masec resolution, and 6 $\sigma$ detections, and assuming
good phase calibration, one could expect parallax measurements
accurate to about 15\% for sources at the Galactic Center distance,
and the accuracy would improve linearly with both signal to noise, and
decreasing distance.

\section{A next stage, more ambitious plan for detecting more sources}

The plan outlined above shows that with 300 hours of ngVLA time, one
would be able to do some important first order work on understanding
the populations of X-ray binaries. It is important to consider,
though, that this survey would discover $\sim100$ BH binaries,
and would only reach Galactic Center distances for the brightest ones.
To build a sample of many hundreds of BH binaries, one could
easily expand this survey through a combination of deeper observations
on the same part of the sky, and coverage of a wider swath of the
Galactic Plane.  Since observations within 1 degree of the Galactic
Plane will reach a height of 500 pc only at the furthest reaches of
the Galaxy, we can expect that increasing the exposure time will
increase the number of sources roughly linearly, since the volume
included will scale as $t^{3/4}$, and until the most numerous source
classes can be reached at the distance of the Galactic Center, the
space density of sources should be rising with increased distance.
Thus to reach a number of $\sim1000$ sources, we could conduct a
survey which detects objects like A0620-00 at the distance of the
Galactic Center, which would require quadrupling the exposure times
for one epoch.



\setcitestyle{numbers}

\end{document}